%% file: paper.tex
\documentclass[conference]{IEEEtran}
\usepackage{amsthm}
\usepackage{graphicx}  
\usepackage{epstopdf}
\usepackage{subfigure}
\usepackage{booktabs}
\usepackage{multirow}
\usepackage{array}
\usepackage{changepage}
\usepackage{listings}
\usepackage{xcolor}
\usepackage{amsmath, amssymb}
\usepackage{algorithm} 
\usepackage{algorithmic} 

\newcommand{\ie}{{\it i.e.,}}

\newcommand{\etc}{{\it etc}}

\begin{document}

%
\title{HBTM: A Heartbeat-based Behavior Detection Mechanism for POSIX Threads and OpenMP Applications}

\author{\IEEEauthorblockN{Weidong Wang$^{1,2}$, Chunhua Liao$^{2}$, Liqiang Wang$^{3}$, Daniel J. Quinlan$^{2}$, Wei Lu$^{1}$\\}
\IEEEauthorblockA{$^{1}$School of Software Engineering, Beijing Jiaotong University, Beijing, China. Email: \{wangweidong,luwei\}@bjtu.edu.cn.\\
$^{2}$Lawrence Livermore National Laboratory, Livermore, CA, USA. Email: liao6@llnl.gov.\\
$^{3}$University of Central Florida, Orlando, FL, USA. Email: lwangcs@gmail.com.  \\
}
}

\maketitle

\IEEEpeerreviewmaketitle

\begin{abstract}

Extreme-scale computing involves hundreds of millions of threads with multi-level parallelism running on large-scale hierarchical and heterogeneous hardware. In POSIX threads and OpenMP applications, some key behaviors occurring in runtime such as thread failure, busy waiting, and exit need to be accurately and timely detected. However, for the most of these applications, there are lack of unified and efficient detection mechanisms to do this. In this paper, a heartbeat-based behavior detection mechanism for POSIX threads (Pthreads) and OpenMP applications (HBTM) is proposed. In the design, two types of implementations are conducted, centralized and decentralized respectively. In both implementations, unified API has been designed to guarantee the generality of the mechanism. Meanwhile, a ring-based detection algorithm is designed to ease the burden of the centra thread at runtime. To evaluate the mechanism, the NAS Parallel Benchmarks (NPB) are used to test the performance of the HBTM. The experimental results show that the HBTM supports detection of behaviors of POSIX threads and OpenMP applications while acquiring a short latency and near 1\% overhead.

\end{abstract}


\IEEEpeerreviewmaketitle

\section{Introduction}
\label{subsec: Intro}

\input{src/Introduction.tex}

The rest of this paper is organized as follows. Section \ref{sec:re} describes an overview of the related work. Section \ref{sec:ftm} describes the heartbeat mechanism. Section \ref{sec:experiments} describes the experimental results. Section \ref{sec:con} concludes the paper and outlines the future work.

\section{Related Work}
\label{sec:re}
\input{src/RelatedWork.tex}

\section{Heartbeat Mechanism}
\label{sec:ftm}

\input{src/Approach.tex}

\section{Experiments}
\label{sec:experiments}
\input{src/Experiments.tex}

\section{Conclusion}
\label{sec:con}
In this paper, the HBTM mechanism for detecting the behaviors of POSIX threads and OpenMP applications was proposed. In the design, two types of implementations were completed (i.e., centralized and decentralized implementations). In both implementations, the unified API was design to guarantee the generality of the mechanism. Meanwhile, the ring-based detection algorithm was designed to ease the burden of the central thread detection at runtime. In addition, the algorithm of the heart rate adjustment was designed to reduce the overhead. To evaluate the mechanism, the NAS benchmarks were used to test the performance of the HBTM. The experimental results show that the HBTM well supports the detection of the behaviors of POSIX threads and OpenMP applications since it can acquire a short latency with near 1\% overhead.

In the future, the complex behaviors in POSIX threads and OpenMP applications will probably be detected using an extended HBTM mechanism. Meanwhile, the extended HBTM will be explored for supporting MPI applications.


\section*{Acknowledgement}

This work is supported in part by NSF under Grant 1118059 and NSFC under Grant 61272353, 61428201. Weidong Wang did this work directed by his mentors, Chunhua Liao and Liqiang Wang, when he was a visiting scholar at the Lawrence Livermore National Laboratory and University of Wyoming, USA.

\end{document}

%% file: src/Introduction.tex
Extreme-scale computing is expected to involve hundreds of millions of processes and/or threads with multi-level parallelism running on large-scale hierarchical and heterogeneous hardware. The behavior of threads could become so important because it may be directly related with the correctness, failure, and resilience of runtime program. POSIX threads (Pthreads) and OpenMP applications  are typically used for shared memory systems. For OpenMP programming, it has drawn a lot of attentions in the HPC community due to its emphasis on structured parallel programming. In order to improve the resilience of the use of these programming models in shared memory systems, the study focuses on detecting the behaviors of application-level multi-threading programs in runtime including busy-waiting, abnormal exit, crash \etc. Each thread's behavior needs to be detected separately. However, detecting these behaviors of pthreads and OpenMP applications is challenging as follows:

1) A good detection mechanism needs to consider the underlying thread model of the application. Current mechanisms may focus on specific multi-threading applications. How to explore the unified interface for different multi-threading applications could be a challenge.

2) Centralized detection method may cause heavy burden on the monitor thread. How to lighten the burden of the monitor thread and minimize impact on the working threads could be another challenge.

3) How to quickly and accurately identify the behavior of any threads in runtime including the status such as busy waiting, abnormal exit, and crash could also be a challenge.

The research of multi-threads' behavior is quite limited even though multi-threading is an indispensable component in high-performance computing and there is less kind of detection tools in POSIX threads and OpenMP applications. In this paper, the goal is to target these above challenges and design heartbeat-based API tools including both centralized and decentralized implementations which can be used in multi-threading applications to detect threads' behavior under low-overhead. Then, the authors use a series of benchmarks including NPB for OpenMP and Jacobi, PI, and matrix multiplication for POSIX threads for executing the designed API tools on NERSC's clusters. The experimental results show that the detection mechanism can acquire a high accuracy of detection with near 1\% overhead.

\subsection{Overview}
\label{subsec: OFTM}

\begin{figure}[bhtp]
  \centering
  \includegraphics[width=7.6cm]{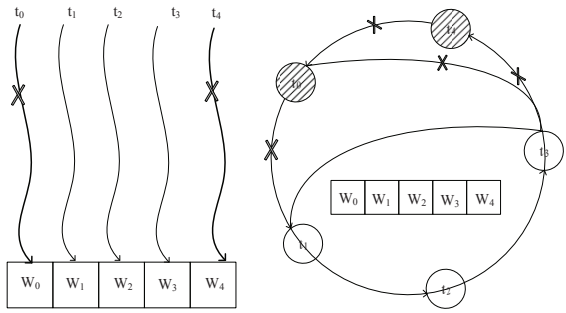}
  \caption{Execution flow of threads' behavior detection.}\label{fig:EMF}
\end{figure}

According to the characters of multi-threading applications and OpenMP applications, both centralized and decentralized implementation of API are designed. For instance, for multi-threading applications, the API of centralized implementation is used and can achieve high performance. While in OpenMP programs, the API of decentralized implementation is easy to use and deploy among slave threads. Figure \ref{fig:EMF} is illustrated to explain both centralized and decentralized mechanisms as follows.

Where $W_n$ denotes a keyword to save the behavior of a thread, and $t_n$ represents a thread. In centralized implementation, a single thread is in charge of the detection work, while in decentralized implementation each thread not only generates heartbeat but also detects its neighbor thread's behaviors. In the right of Figure \ref{fig:EMF}, each thread periodically checks the heartbeat information of its neighbor thread's behavior. Whenever a thread changes its behavior, the current behavior will be detected by one of its neighbors. In the left of Figure \ref{fig:EMF}, the monitor thread periodically checks the status of the working threads.

\subsection{Contributions}
\label{subsec:CA}
The contributions of this paper are shown as follows.

1) The  mechanism proposed is general, which is not dependent on concrete applications. The design of unified API is fully suitable for POSIX threads and OpenMP applications.

2) The decentralized heartbeat mechanism can moderately ease the burden of the monitor thread in a thread team. In the centralized implementation, the number of messages passing is very large between monitor thread and working threads and this may cause a heavy burden on the monitor thread. However, the decentralized heartbeat mechanism can balance the burden among the working threads and achieves a lower overhead compared to the centralized implementation.

3) The mechanism proposed can achieve high accuracy with low overhead under ensuring timeliness. The mechanism automatically adjusts the heart rate to ensure the timeliness and accuracy for detection, meanwhile it can reduce the overhead to a minimum.

%% file: src/RelatedWork.tex
The research of behavior detection in POSIX threads and OpenMP applications refers to judge some threads' status by checking whether their outside behaviors meet the expected ones. Log analysis mechanisms are widely used in behavior detection. Zhou et al. \cite{Zhou2011} propose a model-based mechanism for localizing fault by analysis of runtime data-log during the runtime operation. Cinque et al. \cite{Cinque2010} propose a logging code mechanism to keep track of runtime behaviors. These log-analysis based mechanisms for behavior detection need the extra time cost for computation and may fail to guarantee user's real-time requirements. In order to guarantee real-time requirements, the real-time mechanism such as interface listener is applied to observe the behaviors of a running program. When the behavior is not complied with the intended behavior, the listener will alarm and report the behavior. The interface listener is implemented as detection component, detection service, or detection tool \cite{Delgado2004}. As one of the message passing techniques, the heartbeat detection technique \cite{Lin2011,Li2009,Arora2014} have been widely used in software applications for behavior detection \cite{Hoffmann10,Kumar97,Janssen2010}. Hoffmann et al. \cite{Hoffmann10} present an application-level heartbeat interface. The program can invoke the interface to generate heartbeats, which enable a software system to adapt its behavior to a changing computing environment. In their research, most of work is to improve the software system' s performance using heartbeat technology. Hou et al.\cite{Hou2003} design a distributed heartbeat mechanism to detect server nodes in multi-machine environment, which consists of one master node and multiple standby nodes. 

The limitations of these methods are as follows. 1) Some run-time detection mechanisms rarely consider the problem of high network traffic between detection node and work node. 2) In large-scale parallel computing, the tools of run-time behavior detection for Pthread and OpenMP are more or less missing. Complementary to these mechanisms, a novel heartbeat-based run-time decentralized detection mechanism is proposed for behavior detection of Pthread and OpenMP applications.

%% file: src/Approach.tex
The goal of the heartbeat mechanism is not only to unify the API for supporting POSIX threads and OpenMP applications, but also to ensure the accuracy and timeliness for detection under low overhead.
The mechanism consists of three parts. 1) An application-level unified API is designed and implemented by language ``C''. 2) To lighten the burden of the monitor thread, the task of detection is assigned among each working thread to transform the centralized detection into decentralized detection. 3) To ensure the timeliness and accuracy, the ring-based detection algorithm is designed and deployed among the working threads.

\subsection{Heartbeats API}
\label{sec:HADMA}

Since the heartbeat APIs are meant to be easy to use by programmers, they must be easy to be inserted into POSIX threads and OpenMP applications. The basic heartbeat API is composed of a few functions that can be called from these applications as shown in Table \ref{FUNC}. For each POSIX thread or OpenMP application, heartbeats need to be registered into each thread. The initialization function \emph{Heartbeat\_MultiPthread\_Init} or \emph{Heartbeat\_OpenMP\_Init} is inserted into the application code at the zone of the main thread to initialize heartbeats for each working threads. Each time \emph{Heartbeat\_MultiPthread\_Generate} or \emph{Heartbeat\_OpenMP\_Generate} is called by working threads, a heartbeat event is logged. Each  heartbeat generated is automatically marked with the heartbeat sequence number, current time, and thread ID. Meanwhile, by calling function \emph{Heartbeat\_MultiPthread\_Monitor } or \emph{Heartbeat\_OpenMP\_Monitor}, the monitor thread periodically detects heartbeat sequence numbers and checks whether some heartbeats are recorded or dropped. This can be used to determine the latency between events, and further to judge the behavior of the threads by comparing with the anticipated heartbeat sequence.

\begin{table*}[bhtp]
\centering
\caption{Functions defined in a dynamic pattern.}

\begin{tabular}[width=17cm]{p{8.8cm}p{7cm}}\\
\toprule
Function                                 	                                              & Description  \\
\midrule
int Heartbeat\_MultiPthread\_Init(int Type)                  & Initialize heartbeat in each work thread and a monitor thread using a centralized method, run\_type is option as ``0'' Centralized or ``1'' Decentralized.    \\
void Heartbeat\_MultiPthread\_Generate(int PthreadNum,int LoopNum, int Iteration)                        &Heartbeat generation.   \\
void Heartbeat\_MultiPthread\_Monitor(void)                                        & Detect threads' behaviors.\\
int Heartbeat\_MultiPthread\_Finished(void)                                                             &Terminate heartbeats production and save running records                         \\
int Heartbeat\_OpenMP\_Init(int Type)                       & Initialize heartbeat in each work thread and a monitor thread using a centralized method, run\_type is option as ``0'' Centralized or ``1'' Decentralized.    \\
void Heartbeat\_OpenMP\_Generate(int ThreadNum,int LoopNum, int Iteration)                               &Heartbeat generation.   \\
void Heartbeat\_OpenMP\_Monitor(void)                                               & Detect OpenMP threads' behaviors.\\
int Heartbeat\_OpenMP\_Finished(void)                                                                   & Terminate heartbeats and save records.\\
int Heartbeat\_HeartRate\_Adjust(float Expected\_Heart\_Rate)                                           & Adjust heart rate to balance latency and overhead.\\
\bottomrule
\end{tabular}
\label{FUNC}
\end{table*}

As shown in Figure \ref{fig:IHAO}, four functions including function initialization, function generation ,function monitor, and function finish are inserted into the OpenMP application. The argument ``0'' in \emph{L\_Heartbeat\_OpenMP\_Init(1)}
denotes the centralized implementation is used and the function \emph{omp\_get\_thread\_num()} acquires the OpenMP thread number in a thread team. With the OpenMP program running, the OpenMP working thread generates heartbeats and the monitor thread periodically detects each working threads' behaviors by checking the heartbeat sequences of working threads. Note that ``0'' denotes success and ``1'' denotes failure as the return value in each function.

\begin{figure}[h]
\centering

\lstset{
basicstyle=\tiny,
language=xml,
frame=shadowbox,
rulesepcolor=\color{red!20!green!20!blue!20}
}
\begin{lstlisting}{t}[float=*t]
    ......
    #include "Heartbeat_Support_OpenMP.h"
    #ifndef NUM_THREADS
    #define NUM_THREADS 32
    #endif
    heartbeat_sequence hsh[NUM_THREADS];
    ......
    int main(void){
    int counter=0;
    ......
    Heartbeat_OpenMP_Init(0);
    ......
    omp_set_nested(2);
    #pragma omp parallel sections num_threads(2){
     #pragma omp section{
      Heartbeat_OpenMP_Monitor();
      }
     #pragma omp section{
      #pragma omp parallel{
       #pragma omp for private(j,i) firstprivate(counter)
        for(i=0;i<n;i++)
         for(j=0;j<m;j++){
         ......
         Heartbeat_OpenMP_Generate(omp_get_thread_num(),1,counter);
         counter++;
         }
       }
      }
     }
     Heartbeat_OpenMP_Finished();
     return(0);
    }
\end{lstlisting}
\caption{Inserting Heartbeat API into OpenMP application.}\label{fig:IHAO}
\end{figure}

\subsection{Improvement of Detection Strategy}
\label{sec:Alg}

\begin{figure}[bhtp]
  \centering
  \includegraphics[width=6.2cm]{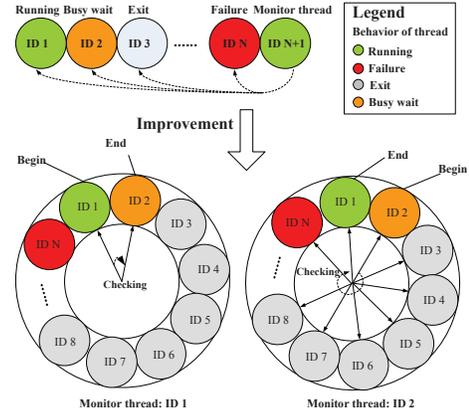}
  \caption{Improvement of detection strategy: decentralized detection.}\label{fig:IDS}
\end{figure}

The goal of the heartbeat detection strategy is to ease the burden of the monitor thread and acquire accurate detection results with a low overhead. Since a great number of messages pass between the working threads and the monitor thread, it may cause heavy burden on the monitor thread. In order to resolve the problem, the ring-based topology is constructed to release the task from the cental monitor thread. So each working thread in ring takes the monitoring task that it periodically checks its neighbor thread's heartbeat sequence to judge its neighbor thread's behavior. In Figure \ref{fig:IDS}, there are $n$ threads ($1, 2, \cdots, n$) in a thread team. In each time, each thread continues to detect its neighbor thread's behavior by checking its heartbeat sequence until it finds a thread alive.

The idea of the algorithm is to judge the behavior of each working thread through its heartbeat sequence. The algorithm is composed of 2 parts. One part refers the detection for alive behavior including either running or busy waiting. The other part means the detection for dead behavior including either failure or exit. Since the monitor task is assigned to each working thread, the following algorithm should be deployed on each thread from the same thread team. The details are in Algorithm \ref{alg:one}. The ``$HSequence$'' denotes a group of heartbeat records for the thread detected and the judgement for threads' behaviors is further discussed in the later section.

\begin{algorithm}[bhtp]
\footnotesize
\caption{Decentralized heartbeat monitor.}
\label{alg:one}
\begin{algorithmic}[1]
\REQUIRE ~~\\
Current Thread $\alpha$'s ID ($ID_{\alpha}$);\\
Thread information global array $T_{ring}$;\\
\ENSURE ~~\\
enum $Behavior=\{exit,failure,busywaiting,alive\}$;\\
\WHILE{(True)}
\STATE   $ID_i$=$T_{ring}.next(ID_{\alpha})$;\\
/*$Part\ 1$: Check clockwise neighbor thread's heartbeat sequence to detect the alive one.*/\;
     \FOR {(Thread $ID_i$ $\in$ $T_{ring}$ in clockwise' direction)}
    \STATE   $HSequence=Get\_Heartbeat\_Sequence$($T_{ring}.next(ID_i)$).
    \IF{IsThreadStart($HSequence$)}
    \STATE {Err=$ThreadIsalive$($HSequence$)}\\
    /*Behavior==1 is a return value that denotes the thread stops running.*/\;
    \IF{Err==1}
       \STATE /*Judge whether the thread has already exited*/\;
         \IF{1==IsExit($HSequence$)}
        \STATE{The thread has already exited.}
        \ELSE
        \STATE{The thread has abnormally terminated.}

        \ENDIF
        \ELSIF{Behavior==0}
       \STATE{}
        /*Behavior==0 is a return value that denotes the thread is running.*/\;
         \IF{1==IsBusywaiting($HSequence$)}
        \STATE{The thread is busy waiting.}

        \ELSE
        \STATE{The thread is running.}

        \ENDIF
  \\  /*Behavior==2 is a return value that denotes the thread failed.*/\;
        \ELSIF{1==IsFailure($HSequence$)}
        \STATE{The thread is failure.}

        \ENDIF
        \ELSE
        \STATE{The thread never got started.}
    \ENDIF
  \\  /*$Part\ 2$: Generate Heartbeat when the application is running.*/\;
            \IF{0==IsApplicationExit()}
            \STATE Save the records into the heartbeat sequences;
            \ELSE
            \STATE Return;
            \ENDIF

           \STATE{Report: thread $ID_i.next$'s behaviors.}
    \ENDFOR
\ENDWHILE\\

\end{algorithmic}
\end{algorithm}

\subsection{Adjustment for accuracy and overhead }
\label{sec:Aco}

The goal of the adjustment is to help multi-threads in making informed decisions when adapting heart rate in the face of a changing overhead from different applications. In this scenario, the heart rate could be adapted to the needs of accuracy of detection with a low overhead. The algorithm consists of two parts. The first part is to detect and calculate current threads' average heart rate. The second part is to update a new heart rate to adapt to the need made by users. The details are shown as shown in Agorithm \ref{alg:two}.

\begin{algorithm}[bhtp]
\footnotesize
\caption{Adjustment of heart rate.}
\label{alg:two}
\begin{algorithmic}[1]
\REQUIRE ~~\\
Current Thread $\alpha$'s ID ($ID_{\alpha}$);\\
Thread information global array $A$;\\
\ENSURE ~~\\
Adjusted Heart rate $Heart\_rate$;\\
/*$Part\ 1$: Calculate current threads' average heart rate.*/\;
\FOR {(Thread $\alpha$ $\in$ $A$)}
\IF{Behavior==$alive$ or Behavior==$busywaiting$}
\STATE{$Sum+$=$A.ID_{\alpha}.HeartRate$;}
\STATE{$Counter++$;}
\ENDIF
\ENDFOR
\IF{$Counter$!=0}
\STATE{$Average\_Heartrate$=$Sum/Counter$;}
\ENDIF\\
/*$Part\ 2$: Heart rate adjustment.*/\;
\IF{($Expected\_Heartrate$-$Threshold$ $\leq$ $Average\_Heartrate$ $\leq$ $Expected\_Heartrate$+$Threshold$)}
\STATE{$Heart\_rate$=Get\_Current\_Heartrate();}
\STATE{Return($Heart\_rate$);}
\ELSE
\STATE{$Time$=(1.0/$Average\_Heartrate$)*($window\_iteration$);}
\STATE{$Amount$=$Time$/(1.0/$Expected\_Heartrate$);}
\STATE{$Iteration$=$window\_iteration$/Amount;}
\ENDIF
\STATE{Heartbeat\_OpenMP\_Generate(32,1,``$Iteration$'');}
\STATE{$Heart\_rate$=Get\_Current\_Heartrate();}
\STATE{Return($Heart\_rate$);}
\end{algorithmic}
\end{algorithm}

The number of heartbeats or window size \ie $window\_iteration$ should be specified to calculate the number of iterations for generating a heartbeat since it is possible for the application to know which window size is most appropriate for the computation it is performing as shown in part three of the algorithm. Users usually try to specify a heart rate for a multi-thread application to adapt to their needs of accuracy and overhead since the high heart rate may take high overhead.

\subsection{Implementation}
\label{sec:IMP}

The implementation of the heartbeats API is provided on Linux by C language and they are useful for POXIS threads or OpenMP programs. The implementation uses files to save heartbeat data and is appropriate for sharing heartbeat information among different threads. In addition, the implementation also uses POSIX shared memory to save the heartbeat information and is appropriate for use among separate threads on the same computer. The implementation adheres to several goals. Firstly, Heartbeat API fully supports POXIS threads and OpenMP applications. Secondly, the decentralized implementation for behavior detection is conducted to ease the burden on the monitor thread. Thirdly, the algorithm of adjusting heart rate is designed to minimize overhead to guarantee the accuracy of the behavior detection.

%% file: src/Experiments.tex

\subsection{Experimental Environment}
\label{sec:EE}

As test inputs, a set of computation kernels are chosen that include Pi iteration, Matrix multiplication, Jacobi iteration for detecting POXIS threads, and NPB benchmark for detecting OpenMP applications. The test platform is Hopper, a Cray XE6 machine provided by the National Energy Research Scientific Computing Center (NERSC) in Livermore. The nodes used run 64-bit SUSE Linux Enterprise Server 11.1 on four quad-core AMD Opteron processors and 129 GB main memory. Each of the AMD processors has 64KB L1 data cache (with a 64-byte line size), 512KB L2 cache, and 6MB L3 cache. The compiler used is GCC 4.8.1.

\subsection{Heartbeats in POSIX thread and OpenMP benchmarks}
\label{sec:EE}

For each benchmark, the authors analyzed the description of the application and chose a loop in the benchmark. Functions corresponding to heartbeats would be inserted in the loop. Table \ref{HIDB} shows where the heartbeat was inserted in terms of the application's needs and the average heart rate that the benchmark achieved over its execution environment. The documentations of benchmarks describe the information for use for each benchmark. By these information, it is simple to find the key loops and insert the call to register a heartbeat in this loop. The total amount of codes required to add heartbeats to each of the benchmarks include four clauses, a header file, and the declaration of a heartbeat data structure. These extra codes can be used to initialize and finalize the heartbeats over the above environment.

\begin{table}[bhtp]
\centering
\caption{Heartbeats in different Benchmarks.}
\begin{tabular}[width=7cm]{p{2.5cm}p{3cm}c}\\
\toprule
Benchmark                      & Heartbeat Location                               &	Heart rate (beats/s)                                  \\
\midrule
   OpenMP NPB-BT                      & Every 1500 iterations                            &    1275.1                                             \\
   OpenMP NPB-CG                      & Every 270000 iterations                          &    1098.2                                             \\
   OpenMP Jacobi               & Every 30000 cycles                               &    205.4                                              \\
   Pthreads PI                & Every 4620000 iterations                         &    10.79                                                  \\
   Pthreads MM       & Every 650 chucks                                 &    120.42                                                  \\
   Pthreads Jacobi         & Every 2300 cycles                                &    1.1                                                 \\
\bottomrule
\end{tabular}
\label{HIDB}
\end{table}

\subsection{Performance of Behaviors}
\label{sec:EE}
\begin{figure}[hbpt]
\centering
\subfigure[Busy waiting]{\includegraphics[width=4.2cm]{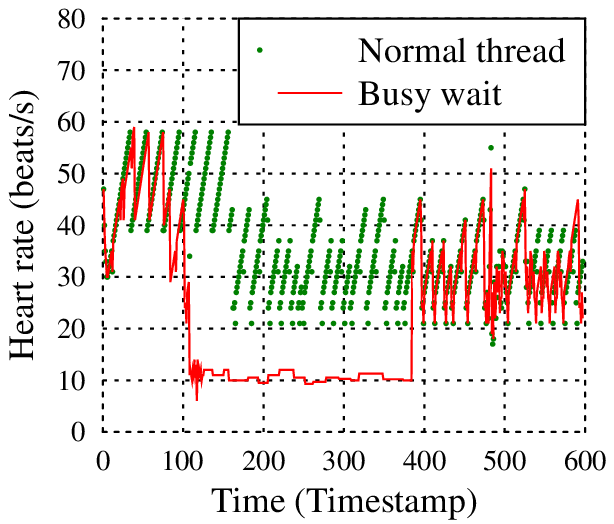}}
\subfigure[Conditional waiting]{\includegraphics[width=4.2cm]{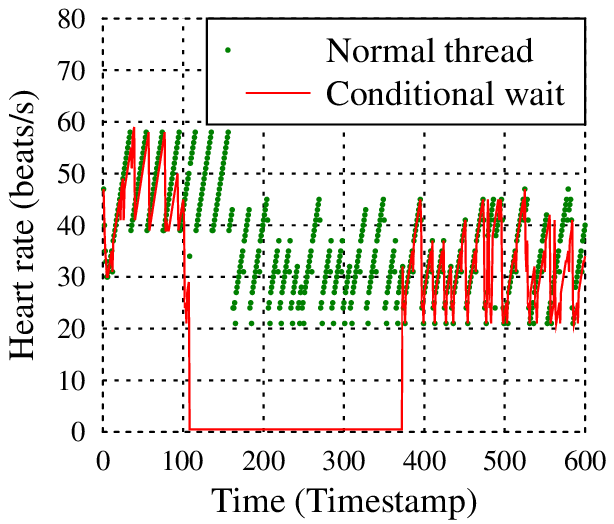}}
\subfigure[Exit]{\includegraphics[width=4.2cm]{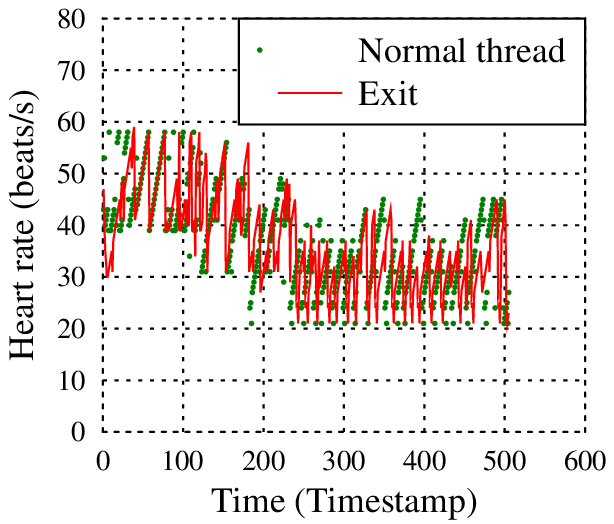}}
\subfigure[Failure]{\includegraphics[width=4.2cm]{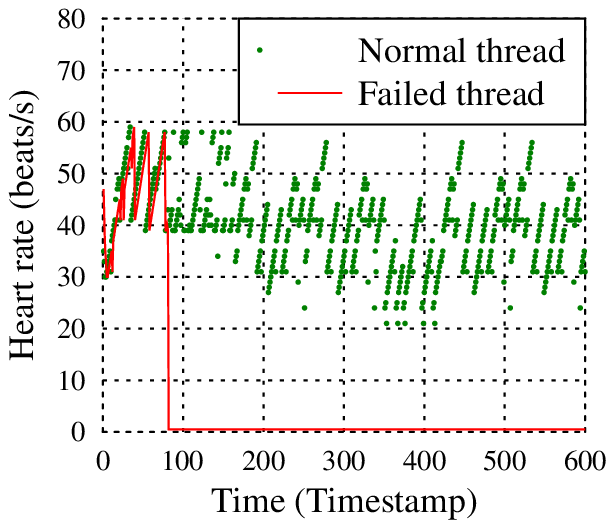}}
\caption{Judgement of behavior based on the heart rate.} \label{fig4}
\end{figure}

The goal of the test is to show the validity of the heartbeat-based behavior detection by analysis of the heart rate from the ``$HSequence$'' mentioned in Algorithm 1. The benchmark POSIX-thread Jacobi was used as the test case. During the execution of the heartbeat-enabled program, the behaviors were manually changed in a specified thread. Then the authors observed the heart rate of the thread and compared it to other normal threads. The experimental results are shown in Figure \ref{fig4}. Let x-axis be the time stamp and y-axis be the heart rate. Figure \ref{fig4}(a) shows that the thread is on the behavior of busy waiting from the time stamp 100 to 400. The heartbeats are generated in different heart rate compared with normal threads. Moreover, the trend for heart rate in this period is smooth and less than the normal thread's heart rate. Therefore, we could judge the behavior is busy waiting. The situation is also applied to Figure \ref{fig4}(b), the only difference is the heart rate (0 beats/s) from the period of conditional waiting. Figure \ref{fig4}(c) shows that the heart rate is normal and both threads almost exit at the same time. Figure \ref{fig4}(d) shows that the thread is on the behavior of failure from the time stamp 80 because the heart rate stopped from that point.

\subsection{Performance Study}
\label{sec:US}

The group of the following three experiments test a series of benchmark NPB (V2.3 C version) and POSIX thread heartbeat-enabled including Pi, Jacobi, matrix multiplication by using up to 8 threads by in GCC 4.8.1 compiler. For this scenario in testing the overhead of HBTM mechanism, the authors run 6 heartbeat-enabled applications using the above benchmarks to record their execution time. The overhead is defined as shown in Equation \ref{formu:3}.

\begin{equation}
\footnotesize
   Overhead=\frac{E_\alpha-E_\beta}{E_\beta} \\
\label{formu:3}
\end{equation}

\begin{figure}[bhtp]
  \centering
  \includegraphics[width=7.5cm]{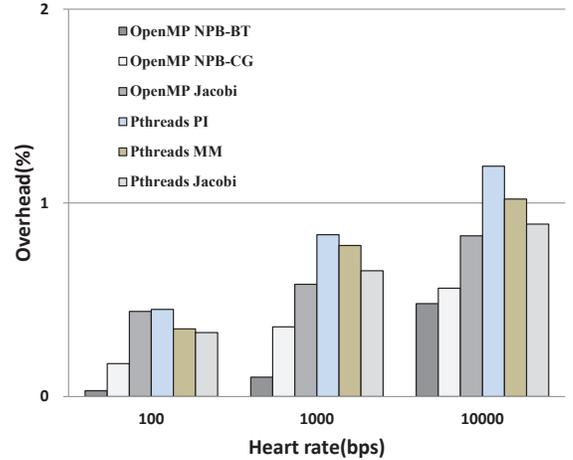}\\
  \caption{Performance on overhead.}\label{fig5}
\end{figure}

Where $E_\alpha$ denotes the execution time of heartbeat-enabled application while $E_\beta$ denotes the execution time of its corresponding benchmark. Figure \ref{fig5} shows the results, the overall overhead is under 1.19\%. The overhead is increasing as the heart rate increases. In addition, the NPB-BT has the best performance of the overhead while the POSIX Pi has the worst performance of the overhead among these benchmarks. Since the heart rate increases, the more heartbeats may take the extra overhead. Meanwhile, the benchmark with the long execution time such as NPB-BT may cost a low overhead because the bigger value of the denominator in Equation (\ref{formu:3}), the smaller result could be generated. Actually, the results indicate that the HBTM mechanism can achieve an almost 1\% overhead by adjusting the heart rate to 1000 beats/s by running Algorithm 2.

\begin{figure}[bhtp]
  \centering
  \includegraphics[width=8.6cm]{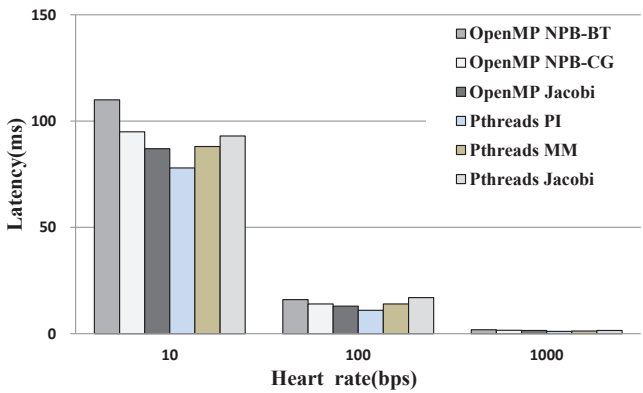}\\
  \caption{Performance on Latency.}\label{fig6}
\end{figure}

For the scenario in testing the latency of HBTM mechanism, the authors executed the same heartbeat-enabled applications and adjusted the heart rate from 10 to 1000 (beats/s). The authors defined the latency that the time it took for a specific heartbeat to rotate around to return the threads' behavior. Figure \ref{fig6} shows the results that the latency is decreasing as the heart rate increases and all the applications get almost the same latency at the same heart rate. This is because the same heart rate determines the latency of the detection and the period of the detection should be shorten by the high heart rate.

\begin{figure}[bhtp]
  \centering
  \includegraphics[width=8.6cm]{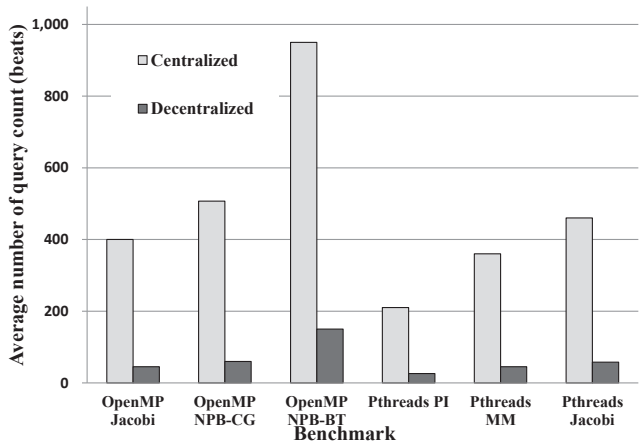}\\
  \caption{Performance on number of query count.}\label{fig7}
\end{figure}

For the scenarios in testing the performance of both centralized and decentralized implementations, the authors run the above 6 heartbeat-enabled applications by inserting either centralized or decentralized API function call, and then recorded the number of messages passing. Figure \ref{fig7} shows the results that in a widow-size of heartbeats, the decentralized implementation achieves by 10\% of the number of the messages passing of the centralized implementation. Since the work of the ring-based heartbeat detection is assigned to each working thread to replace the centra thread's, the decentralized implementation can dramatically ease the burden of cental thread.

\subsection{Performance Comparison}
\label{sec:PC}
For the scenario in comparing the HBTM with other behavior mechanisms, the authors executed the mechanism of the system log analysis proposed by Cinque et al. \cite{Cinque2010}. The results of comparison are shown in Table \ref{tab:Case}. The results consist of three parts. Firstly, the HBTM is fit for detecting 4 behaviors(EX, BW, FA, and CW) while the system log is conducted to detect 2 possible behaviors (EX and FA). Secondly, the HBTM can achieve a less than 2 \textbf{msec} latency while the system log analysis takes more than 2300 \textbf{msec}. Finally, the overhead of the HBTM is almost 1\% while the overhead of the system log analysis is up to 10\%. This is because the system log analysis needs to frequently access the log file that may take a lot of waiting time for IO operations. In addition, retrieving information in log file may be another factor of time-consumption.  Above all, the HBTM achieves the better performance in detecting four specified behaviors.

\begin{table}[htbp]
 \center
 \caption{Evaluation of resilience improvement.}
 {(Note that the notation EX denotes the exit; BW denotes the busy waiting; FA denotes the failure; and CW denotes the conditional waiting.)}
 \begin{tabular}[h]{cccccc}
  \toprule
 \multirow{2}{*}{Performance}  & \multirow{2}{*}{Approach} &  \multicolumn{4}{c}{Behaviors}\\

& &EX  & BW & FA &CW \\

\hline

  \midrule
 \multirow{2}{*}{validity}         & System Log         &  $\surd$    & $-$   &$\surd$       &$-$     \\
                                   & HBTM             &  $\surd$   & $\surd$        & $\surd$      & $\surd$       \\ \hline

 \multirow{2}{*}{Latency(ms)}      & System Log         & 2300       & $-$             & 3020         & $-$     \\
                                   & HBTM             & 1.8        & 1.9             & 1.6          & 1.7    \\ \hline

  \multirow{2}{*}{Overhead(\%)}   & System Log         & 10         & $-$             & 11       & $-$               \\
                                  & HBTM             & 1          & 0.94            & 0.84         & 1.01              \\ \hline

  \bottomrule
 \end{tabular}
 \label{tab:Case}
\end{table}